# Macroscopic and Microscopic Investigation on the History Dependence of the Mechanical Behaviour of Powders

Kadau D*, Brendel L*, Bartels G*, Wolf D E*
Morgeneyer M**, Schwedes J**
*Institute of Physics, University of Duisburg, 47048 Duisburg, Germany
**Institute of Mechanical Process Engineering, Technical University of Braunschweig,
38104 Braunschweig, Germany

As an example for history dependent mechanical behaviour of cohesive powders experiments and computer simulations of uniaxial consolidation are compared. Some samples were precompacted transversally to the consolidation direction and hence had a different history. The experiments were done with two carbonyl iron powders, for which the average particle diameters differed by a factor of ca. 2. Whereas the particle diameter was the only characteristic length in the simulations, the evaluation of the experimental data indicates that at least a second characteristic length must be present.

## 1 Introduction

Powder compaction plays an important role in many areas. Stresses applied to the powder in pharmaceutics and chemical engineering often do not exceed 100 kPa. In these areas, stresses often occur due to the own weight of the powder for example in conveying, or are applied in order to form agglomerates. In civil engineering and powder metallurgy compaction stresses of more than 100 MPa are frequent, for example when metal powders are sintered. Many compaction equations have been proposed, but few of them have a theoretical or even microscopic basis (Denny, 2002). From the theoretical point of view, already the compaction of non-cohesive materials raises fundamental questions like the definition of the hitherto widely used concept of random dense packing (Torquato et al., 2000). Having to take into account cohesion, it becomes clear that the aim of understanding a powder sample's stress-strain behaviour under compaction for the whole range of stresses and strains can only be achieved stepwise using for each stress level the appropriate model and investigation method. In this paper, we focus on the beginning of a uniaxial compaction process and its dependence on the sample's history.

## 2 Methods and Setup

### 2.1 Simulation

An important tool to understand the behaviour microscopically is the modelling on the particle scale, i.e. discrete element methods, in contrast to finite element methods, which are designed for macroscopic description (of a certain experimental setup). So far most computational studies have neglected cohesion between particles. This is justified in dry systems on scales where the cohesive force is weak compared to the gravitational force on the particle, i.e. for sand and coarser material, which can take on densities close to that of random dense packings. Fine powders already show cohesion effects: With decreasing grain diameter cohesive forces lead to highly porous media. For particle diameters in the nanometer range the cohesive force becomes the dominant force, so that particles stick together upon first contact. We use the contact dynamics method which was originally developed in the beginning 1990s for compact and dry systems

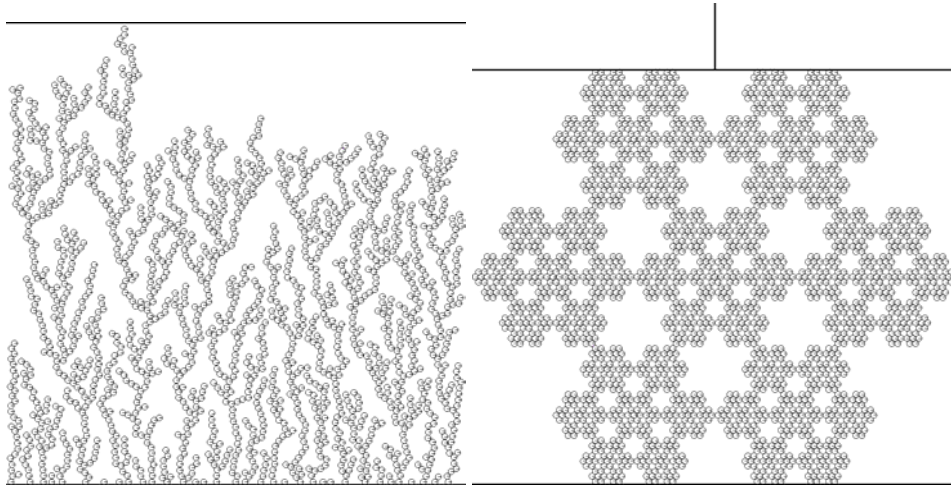

*Figure 1: Ballistic deposit generated with a capture radius of 2.5 particle radii (left) and deterministic fractal consisting of four hierarchies (right)*

with lasting contacts (Jean, 1999, Moreau, 1994). Its concept is based on non-smooth contact laws, that means that steric volume exclusion for perfectly rigid particles and the Coulombian friction law are implemented exactly. We implemented a cohesion model with an attractive force, the cohesion force $F_C$ acting within a finite range (Kadau et al, 2002, Kadau et al, 2003), so that for the opening of a contact a finite energy barrier must be overcome. In addition we implemented rolling friction between two particles in contact, so that large pores can occur within the system (Kadau et al, 2002, Kadau et al, 2003). Using this method we simulate the compaction by an external load in vertical direction of different initial configurations of monodisperse grains: First we use a ballistic deposit (Meakin and Jullien, 1991), built up by ballistic deposition with a capture radius (figure 1, left), where a higher capture radius leads to less dense structures. To check the influence of the precompaction (as done in the experiments) we also did a compaction of a horizontally precompacted ballistic deposit. An approach to low densities other than ballistic deposition is a fractal structure which can be reached by forming each "grain" out of smaller grains in a hierarchical way. One ends up in an ordered fractal with a given number of hierarchies (figure 1, right).

**2.2 Experiment**
*2.2.1 The true biaxial shear tester*
The true biaxial shear tester allows the deformation of a brick shaped bulk solid specimen. The deformation mechanism is composed of a bottom and a top plate which are fixed in a distance of 36,5 mm from each other and four side plates which can be moved independently from a maximum distance of 130 mm to a minimum distance of

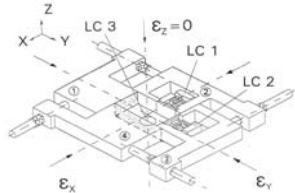

Figure 2: The true biaxial shear tester

70 mm, always staying perpendicular to the adjacent ones (Morgeneyer and Schwedes, 2002). On the specimen's border, the complete stress state is measured with three five component load cells (Schulze et al., 1989), which are installed in the bottom and in two perpendicular load plates (figure 2, LC 1 - 3). For the tests with the true biaxial shear tester, silicon grease is spread on bottom, lateral and top plates. Then they are covered with highly flexible rubber membranes in such a way that the membranes are evenly deformed with the borders of the powder specimen. The friction between specimen and load plates thus is minimised and shear stresses which are worth mentioning compared to the normal stresses cannot be observed. Therefore, the normal stresses on the specimen borders are principal stresses. The principal stress axes coincide with the deformation axes. In this paper we will present experimental data from uniaxial compaction tests. This means, the sample is compacted only in x-direction after sieving and optionally precompacting the powder into the test volume in z-direction.

2.2.2 Carbonyl iron powder

The mechanical properties of several powders have been investigated using the true biaxial shear tester, such as limestone (Nowak, 1993). In this paper, we will show experimental data obtained with carbonyl iron powder (CIP). The misleading prefix "carbonyl" reminds the production of this powder by thermal decomposition of iron pentacarbonyl to pure iron. During the decomposition process, spherical particles develop forming shell-like layers around nuclei. We use a batch of about 2 micron and another of about 4 micron sized primary particles (figure 3).

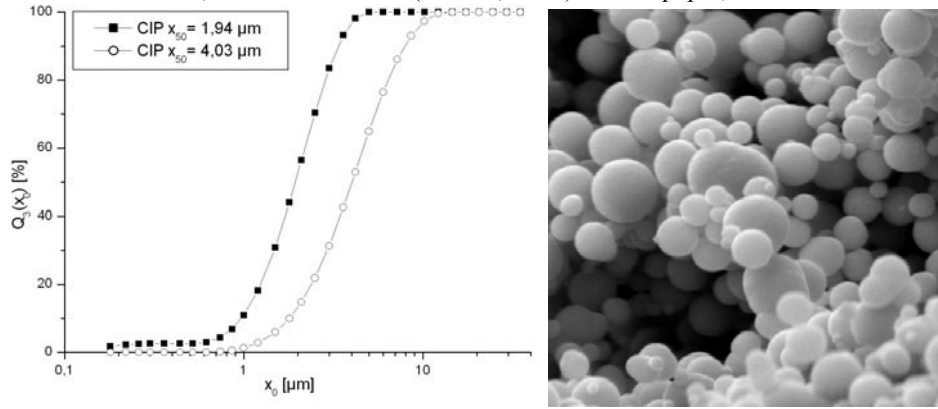

Figure 3: Size distributions (left) and electron microscope image (right) of CIP

# 3 Results

### 3.1 Simulation

The initial configurations in the simulation differ from those in the experiment. Therefore it is important to check in how far this affects the porosity/pressure curves which are our main results. Simulations were done with ballistic deposits of different density as well as with a fractal initial configuration. Figure 4 (left) shows the dependence of the porosity $E$ on the ratio of cohesion force $F_C$ to compacting pressure $\sigma$ for ballistic deposits of different system sizes (but the same initial density). We obtained a data collapse when rescaling the abscissa by the particle radius. Obviously, an initially precompacted configuration has an increased initial density, which survives for low pressures (i.e. high cohesion/pressure ratio) where the initial configuration stays essentially unchanged. For high pressures, the memory of the initial configuration is lost, though, and the curves coincide. This behaviour is found to hold true for other means of initial setup as well (different capture radii, fractal-like structure), which is displayed in figure 4 (right). As the stress becomes strong enough to let large pores collapse, the final solid fraction does no longer depend on the initial configuration, so that the dependence on the stress is the same for all the cases we considered. This is also true for the fractal initial configuration, where the collapse takes place at a rather sharply defined pressure.

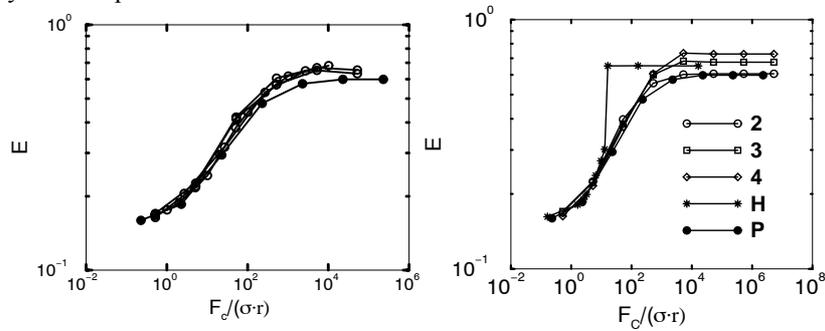

*Figure 4: Left: Porosity as a function of the ratio of cohesion force to (rescaled, two-dimensional) compacting pressure for differently sized ballistic deposits (capture radius = 2.5 particle radii) collapse into one curve (open circles). A lateral precompaction of such a system down to 75% (filled circles) leads to a deviation for high cohesion/pressure ratios only. Right: Porosity of ballistic deposits of different density (using a capture radius of 2, 3, and 4 particle radii) as well as a laterally precompacted ballistic deposit (capture radius 2.5, P) depending on the ratio of cohesion force to (rescaled, two-dimensional) compacting pressure collapse into one curve for low cohesion/pressure ratios only. This is also true for a deterministic fractal (H), albeit with a sharper transition.*

## 3.2 Experiment

Figure 5 (left) shows the experimental results of four compaction tests with the 2 μm powder and three tests with the 4 μm type (to be more precise: having a size distribution median of 1.94 μm and 4.03 μm, respectively). Though the two arrays for the different particle sizes can be clearly distinguished, for one powder type, the curves are not exactly reproducible, but show preparation reminiscences, which reduce merely to an offset in $E$ at high pressures. Therefore, we restrict ourselves to evaluating the derivative of the void ratio $E$ with respect to the pressure, which brings the curves within each array to a collapse (at high pressures). Multiplying furthermore this derivative by the particle radius $r$ (i.e. the distribution's median) yields a collapse of fair quality (for high pressures) as shown in figure 5 (right).

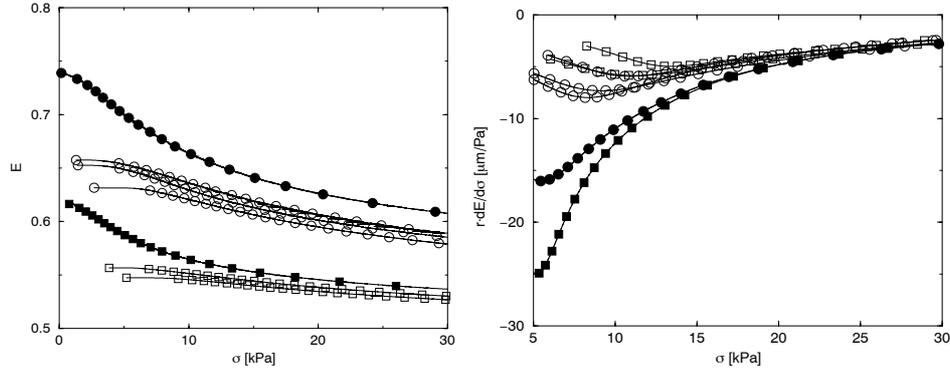

*Figure 5: Left: Void ratio E as a function of acting pressure for smaller (median 1.94 μm, circles) and larger (median 4.03 μm, squares) particles. Filled symbols refer to experiments without a precompaction (cf. section 2.2.1). Right: Correspondence (for high pressures) obtained by differentiating and rescaling.*

## 4 Discussion

The simulations clearly indicate a loss of memory of the initial configuration once a characteristic stress is exceeded. Furthermore it reveals the ratio of cohesion force to particle radius $r$ ($r^2$ in three dimensions) as this characteristic pressure:

$$c = f(\sigma/\sigma_0)$$
with $$\sigma_0 = F_C/r^{d-1}$$

The function $f$ does not depend on system parameters like specimen or particle size, neither cohesion force nor compacting pressure, but it does depend on the initial configuration for large arguments (and, for more general situations, on particle shape and friction coefficient (Kadau et al, 2003). In this sense, it is not universal.

The experimental data, on the other hand, suggest for high pressures a relation of the type:

$$dE/d\sigma = \lambda/(r\,\sigma_l)\,g(\sigma/\sigma_l)$$

where $\lambda$, bearing the dimension of a length, indicates the presence of another characteristic length scale (besides the particle diameter) in the system. It does not depend on the history, which has an influence on the offset (cf. figure 5, left) only.

Without this additional length scale, the $r$-independence of $\sigma_l$ could indicate, taking over the simulation results to three dimensions (i.e. equating $\sigma_0$ and $\sigma_l$), an attractive force of $F_C \propto r^2$, which in turn would suggest $F_C$ being an average magnetic dipole interaction ($F_{dipole} \propto M^2/r^4 \propto r^6/r^4 = r^2$, $M$ being the particle's magnetic moment assumed to be proportional to its volume), not unreasonable for the investigated material.

It must be concluded though, that the available data is at present not sufficient to perform a terminating comparison, since the region of "high pressures" in the experiment spans less than the half of the total range. Furthermore, it is desirable to check for a pressure region where the offset in $E$ (cf. figure 5, left) vanishes. Therefore, an extension of the biaxial shear tester is in preparation.

## 5 Acknowledgements

Useful conversations with Z. Farkas and experimental work by R. Marcos are gratefully acknowledged. We acknowledge support by DFG within SFB 445, "Nano-Particles from the Gas-Phase: Formation, Structure, Properties" and DFG-project "Compaction and Mechanical Properties of Powders".